\documentclass[aps,prl,twocolumn,showpacs,superscriptaddress,10pt]{revtex4-1}
\usepackage{graphicx,amsmath,amssymb,amsfonts,sidecap,color}

\makeatletter
\newcommand*{\rom}[1]{\expandafter\@slowromancap\romannumeral #1@}
\makeatother

\begin{document}

\title{Integer and Fractional Quantum Anomalous  Hall Effect in a Strip of Stripes Model}

\author{Jelena Klinovaja}
\affiliation{Department of Physics, University of Basel,
            Klingelbergstrasse 82, CH-4056 Basel, Switzerland}
\affiliation{Department of Physics, Harvard University,  Cambridge, Massachusetts 02138, USA}
\author{Yaroslav Tserkovnyak}
\affiliation{Department of Physics and Astronomy, University of California, Los Angeles, California 90095, USA}
\author{Daniel Loss}
\affiliation{Department of Physics, University of Basel,
            Klingelbergstrasse 82, CH-4056 Basel, Switzerland}
\date{\today}
	
\begin{abstract}
We study the quantum anomalous Hall effect in a strip of stripes model coupled to a magnetic texture with zero total magnetization and in the presence of strong electron-electron
interactions.
A helical magnetization along the stripes and a spin-selective coupling between the stripes
gives rise to a bulk gap and chiral edge modes. 
Depending on the ratio between the period of the magnetic structure and the Fermi wavelength, the system can exhibit the integer or fractional quantum anomalous Hall effect. In the fractional regime, the quasiparticles have fractional charges and non-trivial Abelian braid  statistics. 

\end{abstract}

\maketitle

{\it Introduction.}
The quantum anomalous Hall effect (QAHE) in two-dimensional systems attracted wide attention recently~\cite{review_AHE,review_AHE_TI}. 
In contrast to the standard quantum Hall effect (QHE)~\cite{Klitzing,Tsui}
induced by a perpendicular uniform magnetic field,
the QAHE can occur at zero total magnetic field, as shown by Pankratov~\cite{Pankratov} and by Haldane~\cite{Haldane} for  models exhibiting a single chiral edge mode, in close analogy to the integer QHE at filling factor $\nu=1$.
The QAHE has been predicted to occur in a variety of materials such as
mercury-based quantum wells~\cite{Zhang,Zhang_2,Wang_2013},
graphene~\cite{Qiao,Mokrousov_2012,Niu_2014}, silicene~\cite{Ezawa_2012} or heavy element based systems~\cite{Garrity_2013}, and
kagome~\cite{Kagome_2011} or optical lattices~\cite{Wu_2008}.
Experimental signatures of the QAHE have recently been reported for magnetically doped topological insulators~\cite{Zhang_3,review_AHE_TI}
and for optical lattices~\cite{ETHZ}.

While the focus of previous work has been on the special case $\nu=1$, we will propose and study here a general model that exhibits not only an arbitrary integer QAHE (IQAHE) with $l$ chiral edge modes, where $l$ is a positive integer, but also a fractional QAHE (FQAHE)  in the presence of strong electron-electron interactions. The latter features edge modes with Abelian quasiparticles that carry fractional charge $e/q$, where $e$ is the elementary electron charge and $q$ a positive odd integer.

Our construction is based on an anisotropic strip of stripes model that allows for treating the electron motion in the direction perpendicular to the stripes as a small perturbation  
\cite{Kane_PRL,Stripes_PRL,Kane_PRB,Stripes_arxiv,yaroslav,Stripes_nuclear,Lebed,Montambaux,Lebed_Gorkov,Yakovenko_PRB,
Wegsheider,Stripe_PRL_exp,Neupert,Oreg}. The strip is embedded in a magnetic material with fixed magnetic texture that couples to the spins of the itinerant electrons.
Aiming at the QAHE, the magnetization is considered to be spatially modulated such that on average it is equal to zero. 
In particular, we assume a helical magnetization inside the stripes. The tunneling between the stripes is spin-selective, which could be achieved by a magnetic medium between the stripes. 
Alternatively,  the same effect can be achieved via spin-orbit interaction and  alternating magnetization instead of  helical fields 
\cite{braunecker_prb_10}.

\begin{figure}[!b]
\includegraphics[width=0.9\linewidth]{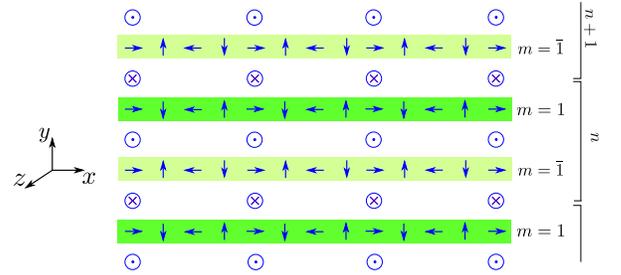}
\caption{A strip of stripes aligned in $y$-direction with the stripes (green  rectangles) lying along  $x$-direction hosting itinerant electrons.  Two neighboring stripes labeled by $m=\pm1$ form a unit cell labeled by  $n$. The localized magnetic moments  inside each stripe 
produce a helical magnetization (blue arrows) with period $\pi l/k_F$ and of opposite chiralities for $m=\pm1$. Here, $k_F$ is the Fermi wavevector inside the stripes. 
The tunneling between stripes is spin-conserving and spin-selective, {\it i.e.,} only spin-up (spin-down) electrons can tunnel between stripes belonging to the same unit cell (to different unit cells). The tunneling amplitude is spatially non-uniform and has a substantial Fourier component at $2k_F/l$.
The total magnetization of the texture is zero.
}
\label{model}
\end{figure}

If the spatial periodic modulation of the magnetic texture is commensurable with the Fermi wavelength, the system is in the QAHE regime if certain conditions are fulfilled. Generally, the strength of the helical fields inside each stripe should dominate over the tunneling amplitude between the stripes. In the fractional regime, electron-electron interactions should be strong enough to enable backscattering of electrons inside the stripes~\cite{Kane_PRL,Kane_PRB,Stripes_PRL}. As we will see, for a fixed chemical potential (fixed magnetic texture), one can tune between integer and fractional QAHE regime by changing the pitch of the magnetic texture (changing the chemical potential).

{\it Model.} We consider a strip of stripes model. The strip aligned in the $y$-direction consists of an array of tunnel-coupled stripes (alternatively referred to as coupled wire construction) aligned in the $x$-direction\  \cite{Stripes_PRL,Stripes_arxiv,Stripes_nuclear,Kane_PRL,Kane_PRB}, see Fig.~\ref{model}.
Two neighboring stripes labeled by indices $m=\pm 1$
form the unit cell.
The electrons propagate freely along the stripes. The tunneling amplitude between two neighboring stripes  is assumed to be weak compared to the Fermi energy inside each stripe such that it can be treated perturbatively. We first treat each stripe as independent and then add the tunneling terms as  small perturbations. Here, we choose the spin quantization axis along the $z$-direction (see below). The chemical potential $\mu$, which sets the Fermi wavevector $k_F$, and the electron density are assumed to be uniform over the entire strip.

The kinetic part of the Hamiltonian corresponding to the $m$th stripe in $n$th unit cell is written as
\begin{align}
H_{0,nm}= \sum_{\sigma=\pm1} \int dx\ \Psi_{n m\sigma}^\dagger  \left(-\frac{\hbar^2\partial_x^2}{2m_0} -\mu\right)\Psi_{n m \sigma} .
\label{kinetic}
\end{align}
Here, the annihilation operator $\Psi_{nm\sigma} (x)$ removes an electron (of effective mass $m_0$ and charge $e$)  with spin $\sigma=\pm 1$  at the position $x$ of the $m$th stripe in $n$th unit cell.

To be more specific, we assume that the effective magnetic field acting on the electron spins inside each stripe rotates in the $xy$-plane as
\begin{align}
{\bf M}_{m}^{(l)}(x) = M [\cos \left(2k_{F}\frac{x}{l}\right) \hat {\bf {x}} + (-1)^{m} \sin (2k_{F}\frac{x}{l})\hat {\bf {y}}],
\end{align}
where $\hat {\bf {x}}$ and $\hat {\bf {y}}$ are unit vectors in $x$- and $y$-direction,  resp.,
$l$ is either a positive integer  or a fraction of the type $1/q$, with $q$ being an odd positive integer (see below). Generally, the magnetization can deviate from above specific form ${\bf M}_{m}^{(l)}(x)$  without changing the main results as long as it has a substantial Fourier component at $2k_{F}/l$.
 The direction of rotation is opposite in the two stripes forming a unit cell. The magnetic spiral is right-handed (left-handed) for stripes with $m=1$ ($m=\bar 1$).
Such a magnetic texture could be produced in several ways. First, it could be generated by extrinsic nanomagnets \cite{braunecker_prb_10,Two_field_Klinovaja_Stano_Loss_2012,helical_graphene,Flensberg,nanomagnets}. 
Second, it could be obtained by making  use of a skyrmion texture 
in the underlying magnetic material \cite{skyrmion_2,skyrmion_1,skyrmion_4,skyrmion_tokura,Tserkovnyak_2014}.
Third, one can imagine a helix of local magnetic moments, such as nuclear spins or magnetic impurities, formed intrinsically via, for example, RKKY interaction in an underlying 
strip of stripes~\cite{
RKKY_Basel,RKKY_Simon,RKKY_Franz,Daniel_RKKY,
Exp_RKKY,Stripes_nuclear}. 
Finally, we assume that the coupling of the texture field to the electron orbit is negligible.

The corresponding Zeeman term is given by
\begin{align}
H_{M,nm}^{(l)}=\sum_{\sigma,\sigma'} \int dx\  \Psi_{n m\sigma}^\dagger 
\Big ( \mu {\bf M}_{m}^{(l)} \cdot \boldsymbol \sigma \Big )_{\sigma\sigma'}\Psi_{n m\sigma'} ,
\label{Zeeman}
\end{align}
where 
$\boldsymbol \sigma = (\sigma_x,\sigma_y,\sigma_z)$ is a vector composed of the Pauli matrices $\sigma_i$ representing the electron spin. 
The energy scale given by $\Delta_M = \mu M$ depends on the coupling constant $\mu$ between the local magnetization $M$ and the electron spin.

The aforementioned tunneling between two neighboring stripes is assumed to be both spin and position dependent. This can be achieved, for example, by placing nanomagnets that polarize the medium between the stripes or be a consequence of intrinsic magnetic ordering.
 As a result, the hopping between stripes inside the same unit cell (belonging to neighboring unit cells) is allowed only for spin-up (spin-down) electrons.
In addition, we assume that the magnitude of the tunneling amplitude $t_{(l)}(x)=t_0+2t_l\cos(2k_F x/l)$  [with $t_0$ and $t_l$, for simplicity, being non-negative] is spatially modulated, giving
\begin{align}
&H_{t} ^{(l)}= \sum_{n,\sigma} \int dx\ t_{(l)}
\label{tunneling} \Big( \Psi_{n1\sigma}^\dagger   \Psi_{[n+(1-\sigma)/2]\bar 1\sigma}   + H.c.\Big).
\end{align}
Here, we consistently choose the spatial modulations of the tunneling term $t_{(l)}(x)$ and the magnetic texture ${\bf M}_{m}^{(l)} (x)$ to be the same as it is the case for both extrinsically and intrinsically imposed periodicity.

{\it Single chiral  edge mode.}
We begin with the single edge mode regime characterized by the effective `filling factor' $ l=1$. 
First we consider the case with chiral edge modes propagating along $y$-direction and then along $x$-direction.

For the moment, we assume that the system is periodic in the $y$-direction and contains $N$ unit cells, {\it i.e.,} $2N$ stripes, and search for edge modes that are localized in the $x$-direction and propagate along the $y$-direction. For convenience we introduce the momentum $k_y$  defined by the Fourier transform $\Psi_{k_ym\sigma}(x) = \sum_{n} e^{ink_y a_y}  \Psi_{n m \sigma}(x)/\sqrt{N}$, where $a_y$ is the unit cell size. By analogy, we introduce the momentum $k$ in the $x$-direction taken from the Fermi wavevector $k_F$ \cite{MF_PRB}. 
The topological phase transition is determined by the conditions under which the topological gap closes \cite{Appendix}.  Excluding the trivial case $\Delta_M=0$, we find that the spectrum has a gap at the chemical potential, except for these special parameter values $k=k_y=0$ and $t_1^2-t_0^2 = \pm \Delta_M t_1$, in the vicinity of which the bulk gap closes and reopens as function of these parameters \cite{Appendix}. This behavior suggests that there is a  topological phase transition, which, however, needs to be checked in more detail.
 
The general expression for the energy spectrum is too involved if all parameters are non-zero. 
Thus, we focus on the case $t_0=0$ (see also below), which we can treat  fully analytically. The bulk spectrum is given by
\begin{align}
&E_{0,\pm} = \pm \sqrt{(\hbar \upsilon_F k)^2 + t_1^2},\\
&E_{\pm,\pm} = \pm \sqrt{(\hbar \upsilon_F k)^2 + t_1^2 +\Delta_M ^2 \pm 2 t_1 \Delta_M \cos(k_ya_y/2)}, \nonumber
\end{align}
where  $E_{0,\pm}$ is twofold degenerate. 
The energy of the edge modes is found to be $\epsilon=t_1 \sin(k_ya_y/2)$ with $k_ya_y\in(-\pi,\pi)$
under the condition $\Delta_M> t_1 \cos(k_ya_y/2)$ (see Ref. \onlinecite{Appendix} for more details). 
The localization length is determined as $\xi={\rm max}\{\xi_0,\xi_-\}$ with $\xi_0=\hbar\upsilon_F/[t_1 \cos(k_ya_y/2)]$ and $\xi_-=\hbar\upsilon_F/[\Delta_M-t_1 \cos(k_ya_y/2)]$.

\begin{figure}[!b]
\includegraphics[width=0.65\linewidth]{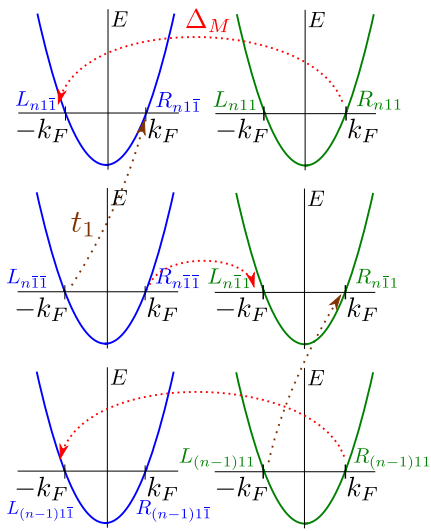}
\caption{The energy spectrum of a strip of stripes in the momentum representation for  $l=1$. Each parabola, being initially twofold degenerate in spin, represents the spectrum of a particle freely propagating in $x$-direction. The magnetic  texture opens partial gaps of the size $2\Delta_M$ in the bulk spectrum but leaves gapless  modes
$L_{n 11} (x)$, $R_{n 1\bar 1} (x)$, $L_{n \bar 1 \bar 1}(x)$, and $ R_{n \bar 1 1} (x)$, with $n$ being the unit cell index. The spin-selective hopping $t_1$ between stripes gaps out these gapless modes in the bulk. However, there remain still gapless propagating modes at the edges: one chiral mode $R_{N 1\bar 1} (x)$ at the upper edge and one chiral mode $L_{1 \bar 1\bar 1} (x)$ at the lower edge.
}
\label{nu1}
\end{figure}

The presence of edge modes in $x$-direction can be easily seen in the limit
$\Delta_M \gg t_0,t_1$ (deep inside the topological phase), where we can use a two-step perturbation procedure. In a first step, we rewrite the Hamiltonian consisting of the kinetic [$H_0$, see Eq. (\ref{kinetic})] and Zeeman  [$H_M^{(1)}$, see Eq. (\ref{Zeeman})] terms in a basis composed of slowly-varying left $L_{nm\sigma} (x)$ and right $R_{nm\sigma} (x)$  movers inside each stripe \cite{MF_PRB},
$\Psi_{nm\sigma}(x)= e^{ik_Fx} R_{nm\sigma} (x) +e^{-ik_Fx}L_{nm\sigma} (x)$.
In what follows, we define the Hamiltonian density $\mathcal H$ as $H = \sum_{n=1}^{N} \int dx\ \mathcal H(x)$. As a result of such linearization, the kinetic term assumes the form
\begin{align}
&{\mathcal H}_{0}= \sum_{m,\sigma=\pm1} i\hbar \upsilon_F   \Big[L^\dagger_{nm\sigma} \partial_x   L_{nm\sigma} - R^\dagger_{nm\sigma} \partial_x   R_{nm\sigma}    \Big],
\end{align}
while the Zeeman term becomes
\begin{align}
&{\mathcal H}_{M}^{(1)}=\Delta_M  \big[ R_{n 11}^\dagger 
L_{n 1\bar 1} +R_{n \bar 1 \bar 1}^\dagger L_{n \bar 1 1}  + H.c.\big].
\end{align}
Here, we neglect all fast-oscillating terms.
We note that $H_M^{(1)}$ gaps out only  half of the modes \cite{braunecker_prb_10}.  The modes $L_{n 11}$, $R_{n 1\bar 1}$, $L_{n \bar 1 \bar 1}$, and $ R_{n \bar 1 1} $ do not occur in $H_M^{(1)}$ and thus are still  gapless, see Fig. \ref{nu1}. 

In a second step, we include the tunneling term [see Eq. (\ref{tunneling})]. Here, we note that a momentum-conserving tunneling process of amplitude $t_0$ 
does not directly couple the above-mentioned gapless modes and, thus, is irrelevant in leading order of  perturbation theory. (We note here that similarly we can neglect
other possible tunneling terms, not considered above, if they are non-resonant or are of smaller magnitude than $t_1$ and thus are not strong enough to close the bulk gap.)
 In contrast, the $t_1$ tunneling terms connect the gapless modes not affected  by the Zeeman term, 
\begin{align}
{\mathcal H}_{t}^{(1)} = t_1   [L_{n11}^\dagger  R_{n\bar 11}  + R_{n 1\bar1}^\dagger  L_{(n+1) \bar 1\bar 1} + H.c.].
\end{align}
This coupling results in a fully gapped bulk spectrum. However, there is one gapless mode left at each of the two edges of the strip. In the unit cell $n=1$ ($n=N$), the mode $L_{1 \bar 1\bar 1}$ ($R_{N 1\bar1} $) is gapless. 

The  single chiral edge modes found above correspond to the QAHE edge states as the magnetic field is zero on average. The direction of propagation of the modes is set by the helicity of the magnetic texture in the bulk.
The spin polarization of the edge modes depends on whether the corresponding boundary stripe is $m=1$ or $m=\bar 1$. We note that the scheme described above can be easily generalized to other integer values of $l$, resulting in  $l$ propagating modes at the same edge \cite{Appendix}.

{\it  Fractional quantum anomalous Hall effect.}
Next, we focus on the fractional QAHE characterized by $l=1/q$, with $q$  a positive odd integer. This regime can be obtained by choosing $t_{(1/q)} (x)$ and ${\bf M}_{(1/q)}(x)$
such that the new Fermi wavelength  $2\pi/ k_F$ is a multiple of the scattering term periodicity $ \pi q / k_F$. 
Under this condition, however, the direct scattering between right- and left-movers is not possible due to  momentum conservation. As a consequence, the opening of gaps can then occur  only in the regime of strong electron-electron interactions where backscattering, which can compensate for the momentum mismatch, plays a crucial role \cite{Kane_PRL,Kane_PRB,Stripes_PRL,PF_1}.
Below we focus  on the regime of $l=1/3$, but the results can be straightforwardly generalized to any `filling factor' of the form $l=1/q$.

\begin{figure}[!bt]
\includegraphics[width=0.65\linewidth]{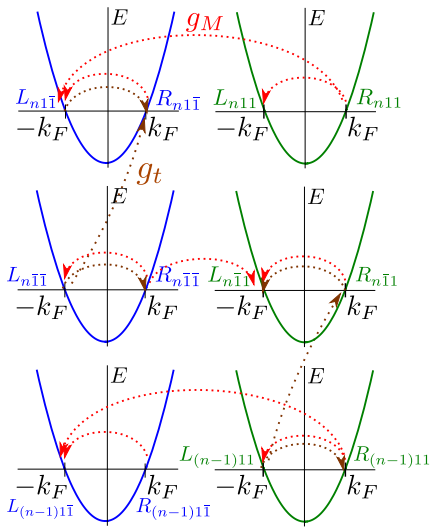}
\caption{The scattering processes for a strip of stripe model at the filling factor $l=1/3$ which include strong interactions. The spectrum of spin-up (right parabola, green) and spin-down (left parabola, blue) electrons are shown separately for each stripe. The helical magnetic field of period $\pi/3k_F$ can open a gap in the spectrum only if spin-conserving backscattering  processes due to electron-electron interactions is taken into account (red dotted lines). The same applies to the tunneling terms (brown dotted lines).}
\label{AFQHE13}
\end{figure}

To begin with, we make use of backscattering terms \cite{giamarchi_book} due to electron interactions (see Fig.~\ref{AFQHE13}) and construct the spin helix scattering term ${ H}_{M}^{(1/3)}$ in leading order that conserves momentum,
${\mathcal H}_{M}^{(1/3)}=g_M 
 \Big( R_{n 11}^\dagger 
L_{n 1\bar 1} [R_{n 11}^\dagger 
L_{n 1 1} ][R_{n 1 \bar 1}^\dagger 
L_{n 1\bar 1} ] +R_{n \bar 1 \bar 1}^\dagger  L_{n \bar 1 1} 
[R_{n \bar 1 \bar 1}^\dagger  L_{n \bar 1 \bar1} ] [R_{n \bar 1  1}^\dagger  L_{n \bar 1 1} ]\Big) + H.c.
$
Similarly, the tunneling term ${H}_{t}^{(1/3)}$ in leading order in the interactions, which conserves both momentum and spin, is given by
${\mathcal H}_{t}^{(1/3)} =  g_t 
  \Big(L_{n11}^\dagger  R_{n\bar 11}   [L_{n11}^\dagger   R_{n11}][L_{n\bar 11}^\dagger   R_{n\bar 11}] + R_{n 1\bar1}^\dagger L_{(n+1) \bar 1\bar 1} [R_{n 1\bar1}^\dagger (x) L_{n \bar 1\bar 1} ][R_{(n+1) \bar 1\bar1}^\dagger  L_{(n+1) \bar 1\bar 1} ]+ H.c.\Big).$
Here, $g_t$ ($g_M$) is proportional to the initial tunneling amplitude $t_{1/3}$ ($\Delta_M$) and to $g_B$, where $g_B^2$ describes the strength of the backscattering term arising from electron-electron interactions. Importantly, the term $H_{M}^{(1/3)}$ commutes with $H_{t}^{(1/3)} $, and, thus, they can be ordered simultaneously in the RG sense (see below) \cite{yaroslav,Neupert,Oreg}.

To treat electron-electron interactions in the one-dimensional stripes, we make use of the  Luttinger liquid formalism based on bosonization techniques \cite{giamarchi_book}. First, we introduce chiral fields $\phi_{rnm\sigma}$  via
$R_{nm\sigma}(x) = e^{i \phi_{1nm\sigma}(x)}$ and $L_{nm\sigma}(x) = e^{i \phi_{\bar 1nm\sigma}(x)}$.
We do not take into account Klein factors explicitly in the present work. However, by choosing the commutation relation for bosonic fields as
$[\phi_{rnm\sigma}(x),\phi_{rnm\sigma}(x')]=ir\pi\ {\rm sgn}(x-x')$, the anticommutation relation for the same fermion branch can be satisfied explicitly. At the same time, following standard practice, we neglect the anticommutation relations between two different fermion branches to keep calculations simple \cite{yaroslav}.

To progress further, we introduce new bosonic fields $\tilde\phi_{rnm\sigma} =(2\phi_{rnm\sigma}-\phi_{\bar rnm\sigma})/3$, which
 obey non-standard commutation relations,
$[\tilde\phi_{rnm\sigma}(x),\tilde\phi_{rnm\sigma}(x')]=(ir\pi/3)\ {\rm sgn}(x-x')$.
This leads to a simplified form of the nonlinear terms in the Hamiltonian. In particular, $H_{M}^{(1/3)}$ and $H_{t}^{(1/3)}$ read
\begin{align}
&{\mathcal H}_{M}^{(1/3)}= g_M  \sum_{\sigma=\pm 1}  \cos [3(\tilde \phi_{\bar 1 n \sigma \bar \sigma}-\tilde \phi_{ 1 n \sigma \sigma})] ,
\nonumber\\
&{\mathcal H}_{t}^{(1/3)}= g_t  \sum_{\sigma=\pm 1}  \cos [3(\tilde \phi_{\bar 1 n \sigma \sigma}-\tilde \phi_{ 1 [n+(1-\sigma)/2] \bar \sigma \sigma}) ]. 
\end{align}
In the strong coupling regime, the cosines are getting pinned (ordered) in such way that   the total energy is minimal \cite{yaroslav,Neupert,Oreg,David,Mong,Stern_PF,cheng_PF,PF_TI}. This means that all fields except for $\tilde \phi_{ 1 1  1 \bar 1}$ in the first unit cell and $\tilde \phi_{\bar 1 N \bar 1  \bar 1}$ in the last unit cell are gapped. These remaining gapless modes represent the edge modes. The elementary excitations of the edge modes are non-trivial, as follows directly from the form of ${\mathcal H}_{M,t}^{(1/3)}$ and the commutation relations between the $\tilde \phi$-fields derived above: they carry a fractional charge $e/3$ with corresponding Abelian braid statistics~\cite{Kane_PRL,Oreg}.
Hence, the presence of the edge modes and their properties confirm that the system is in the FQAHE regime corresponding to an effective  `filling factor' $l=1/3$.

{\it Alternative schemes with spin-orbit interaction.} We note that the model considered above is mathematically equivalent to the one based on the combination of  spin-orbit interaction (SOI) and uniform magnetic fields instead of the helical fields inside the stripes \cite{braunecker_prb_10}.
The SOI with wavevector $k_{so}$ polarizes spins along the $z$-axis and has opposite signs for $m=\pm1$ stripes. The local uniform field acting inside the stripes is applied perpendicular to the $z$-axis, let say, in the $x$-direction, and also has opposite signs for $m=\pm1$ stripes such that no total magnetization is created.
The tunneling is again spin-selective as before but is now modified with the period $\pi/2k_{so}$. In this configuration, all scattering terms written in the basis of right and left movers are the same as in Fig. \ref{nu1}. Here, the chemical potential is tuned such that the Fermi wavevectors satisfy $k_{F\pm}=k_{so}(1 \pm l)$.

{\it Conclusions and Outlook.} We constructed an anisotropic two-dimensional model  with spatially periodic magnetization which exhibits both IQAHE and FQAHE. In particular, in the FQAHE regime, the system possesses fractional charges $e/q$, where $q$ is an odd positive integer. In addition, the model could be generalized to superconducting heterostructures hosting
such exotic particles as Majorana fermions or parafermions 
\cite{cheng_PF,Stern_PF,Barkeshli,Vaezi,David,Mong,
PF_1,PF_2,Vaezi_2,PF_TI}.
The advantage of the QAHE in this context over the standard QHE  lies in the zero  total magnetic field needed for the former, and, consequently, in a potentially much larger proximity-induced pairing gap. Moreover, the strip of stripes model could, in principle,  be extended to describe other QAHE analogs of the QHE at the filling factors $\nu=r/q$ (where both $r$ and $q$ are positive integers, and $q$ is odd) as well as at even denominator filling factors which could host non-Abelian quasiparticles.

{\it Acknowledgments.}
We acknowledge support from the Harvard Quantum Optical Center, the Swiss NSF,  NCCR QSIT, and
funding from FAME (an SRC STARnet center sponsored by
MARCO and DARPA).


\newpage

\section{Supplemental Material for "Integer and Fractional Quantum Anomalous  Hall Effect in a Strip of Stripes Model}

\subsection{Edge states in $y$-direction for $l=1$}

We assume that the system is periodic in the $y$-direction and contains $N$ unit cells, {\it i.e.,} $2N$ stripes, and search for edge modes that are localized in the $x$-direction and propagate along the $y$-direction \cite{Stripes_PRL_A,Stripes_arxiv_A}. For convenience we introduce the momentum $k_y$  defined by the Fourier transformation
$\Psi_{k_ym\sigma}(x) = \frac{1}{\sqrt{N}}\sum_{n} e^{ink_y a_y}  \Psi_{n m \sigma}(x)$.
In addition, the annihilation operator $\Psi_{k_ym\sigma}$ can be represented in terms of slowly-varying right-mover $R_{k_y m \sigma}(x)$  and left-mover $L_{k_y m \sigma}(x)$  fields defined close to the Fermi points $\pm k_F$ as $\Psi_{k_ym\sigma}(x) = R_{k_ym\sigma}(x) e^{ik_{F}x} + L_{k_ym\sigma}(x) e^{-ik_{F}x}$ 
\cite{MF_PRB_A,Two_field_Klinovaja_Stano_Loss_2012_A,giamarchi_book_A}.
The total Hamiltonian is diagonal in momentum
$H=\sum_{k_y} H_{k_y}$ and,
for our convenience, 
can be expressed in terms of the associated Hamiltonian density ${\cal H}_{k_y}(x)$, via $H_{k_y}=\int dx\ \Psi_{k_y}^\dagger(x) {\cal H}_{k_y} \Psi_{k_y}(x)$, where
\begin{widetext}
\begin{align}
&{\cal H}_{k_y}=\hbar \upsilon_F \hat k \lambda_3 + \Delta_M \lambda_1 \sigma_1/2 -
 \Delta_M \lambda_2 \sigma_2 \eta_3/2 + t_0 [1+\cos(k_ya_y)]\eta_1/2 +t_0 [1-\cos(k_ya_y)]\sigma_3 \eta_1 /2 \nonumber \\
&\ \ \ \ \ \ \ + t_1 [1+\cos(k_ya_y)]\lambda_1\eta_1/2 +t_1 [1-\cos(k_ya_y)]\lambda_1 \sigma_3 \eta_1/2 .
\end{align} 
\end{widetext}
Here, we choose the basis $\Psi_{k_y}(x)$=($R_{k_y,11}(x)$, $L_{k_y,11}(x)$, $R_{k_y,1\bar 1}(x)$, $L_{k_y,1\bar 1}(x)$, 
$R_{k_y,\bar11}(x)$, $L_{k_y,\bar11}(x)$, $R_{k_y,\bar1\bar 1}(x)$, $L_{k_y,\bar 1\bar 1}(x)$) composed of the right- and left-movers. The momentum operator $\hat k= -i \partial_x$ is determined close to the Fermi points $\pm k_F$ and $\upsilon_F=\hbar k_F/m_0$ is the Fermi velocity. The Pauli matrix $\lambda_i$ ($\sigma_i$) acts on right/left-mover (spin) space and $\eta_i$ act on sublattice space with $i=1,2,3$.
 
\begin{figure}[!t]
\includegraphics[width=0.85\linewidth]{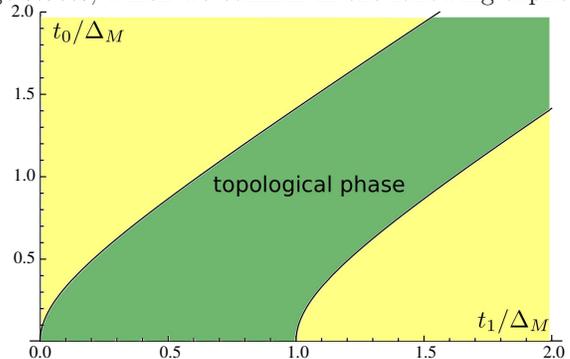}
\caption{The topological phase diagram as a function of the system parameters $t_0/\Delta_M$ and $t_1/\Delta_M$, see Eq. (\ref{TC}).
The topological phase (green area) hosts edge modes. The topologically trivial phase (yellow area) corresponds to the fully gapped system without any state inside the bulk gap.
}
\label{TI}
\end{figure}

To determine if the system is gapped or not at the Fermi level, it is sufficient to look at  zeroes of the determinant of ${\cal H}_{k_y}$,
\begin{align}
&{\rm det} {\cal H}_{k_y} =\Delta_M^4 t_1^2 \Big( t_1^2\sin^2(k_y a_y)+2 (\hbar\upsilon_F k) ^2 [1+\cos(k_y a_y)]\Big) \nonumber\\
&+\Big([(\hbar\upsilon_F k)^2+t_1^2-t_0^2]^2 + \Delta_M ^2 
[(\hbar\upsilon_F k)^2-t_1^2 \cos(k_y a_y)]\Big)^2.
\end{align}
We note that the effective Hamiltonian ${\cal H}_{k_y}$ possesses particle-hole symmetry.
As a consequence, $U^\dagger_P  {\cal H}_{k_y}(k) U_P = -{\cal H}_{k_y}(k)$, where $U_P= \lambda_1\eta_2\sigma_3 K$, and $K$  denotes complex conjugation acting as $K\Phi=\Phi^*$. Thus, the energy spectrum is symmetric around zero, {\it i.e.} both $E$ and $-E$ are eigenvalues of the Hamiltonian.
The system is gapless in two cases. The first  case is trivial and corresponds to 
$\Delta_M=0$ and $\hbar\upsilon_F k = \pm \sqrt{t_0^2 -t_1^2}$ under the condition that $t_0>t_1$. The second case corresponds to $k=k_y=0$ and 
\begin{align}
 t_1^2-t_0^2 = \pm \Delta_M t_1. 
 \label{TC}
\end{align}
In this case, which is the main focus of this work,
the spectrum can be seen to have a gap except at those special parameter values where the gap closes and reopens as function of these parameters, see Fig. \ref{TI}. This behavior suggests that there is a  topological phase transition with edge states, which we confirm in the following explicitly.

\subsection{Edge state wavefunctions for $l=1$ and $t_0=0$}

In this section we derive the edge modes wavefunction for the special case $t_0=0$ and $\Delta_M>t_1$ such that the system is in the topological phase, see Fig. \ref{TI}.

We impose vanishing boundary conditions on the wavefunctions at the left and right ends of each stripe. 
For example,  wavefunctions should go to zero at the left end of each strip, {\it i.e.}, $\Phi(x=0)=0$.
The energy of the bound state is found to be $\epsilon=t_1 \sin(k_ya_y/2)$ with $k_ya_y\in(-\pi,\pi)$
under the condition $\Delta_M> t_1 \cos(k_ya_y/2)$.
The corresponding wavefunction describing the edge states is written in the basis $(\Psi_{k_y 11},\Psi_{k_y 1 \bar1},\Psi_{k_y \bar 11},\Psi_{k_y \bar 1 \bar1})$ as
\begin{widetext}
\begin{align}
\Phi(x)=
\begin{pmatrix}
e^{-i(k_Fx+ k_y a_y/2)}\\
-i e^{-i(k_F x - k_y a_y/2)}\\
i  e^{-ik_F x}\\
-  e^{ik_F x}
\end{pmatrix}e^{-x/\xi_0}
+\begin{pmatrix}
-e^{-i(k_Fx-k_y a_y/2)}\\
i e^{i(k_F x+k_y a_y/2)}\\
-i  e^{ik_F x}\\
e^{-ik_F x}
\end{pmatrix}e^{-x/\xi_-},
\end{align}
\end{widetext}
where the localization lengths are given by
$\xi_0=\hbar\upsilon_F/[t_1 \cos(k_ya_y/2)]$ and $\xi_-=\hbar\upsilon_F/[\Delta_M-t_1 \cos(k_ya_y/2)]$.

\begin{figure}[!bt]
\includegraphics[width=0.65\linewidth]{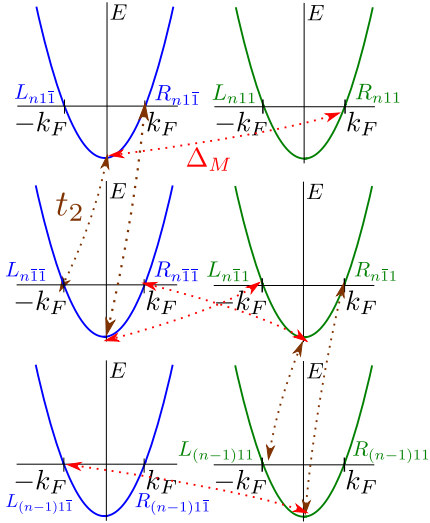}
\caption{The energy spectrum of a strip of stripes in the momentum representation for the  IQAHE regime with $l=2$. The notations are the same as in Fig. 2 of the main text.
The magnetic skyrmion texture combined with spin-selective tunneling terms (both have a period $\pi/2 k_F$) opens partial gaps of the size $\Delta_M t_2/E_F$ in the bulk spectrum but leaves two gapless  modes at each edge of the stripe. 
}
\label{nu2}
\end{figure}

\subsection{Multiple edge modes}

We consider the IQAHE  for higher effective `filling factors' $l$ (with $l$ being a positive integer). 
We focus on multiple ($l$) chiral edge modes propagating in the same direction along the $x$ axis as they are easier to derive, using a  perturbative approach, in the limit of $H_M^{(l)}$ [see Eq. (3) in the main text] being dominant. 

A magnetic helix with period $\pi/l k_F$ [$H_M^{(l)}$, see Eq. (3) in the main text] opens gaps in the spectrum only in combination with the spin-selective hopping term between stripes modulated with the period $\pi/l k_F$. Below, we just focus on the example for $l=2$ (see Fig. \ref{nu2}), but the same method can be easily generalized to other positive integer values of $l$.

As a consequence, the effective coupling between the right ($R_{n 1 \sigma}$) and left ($L_{n \bar 1 \bar \sigma}$) movers [between the right ($R_{n \bar 1 \sigma}$) and left ($L_{(n-1)  1 \bar \sigma}$) movers]  is determined in second-order perturbation expansion \cite{Stripes_arxiv_A,Stripes_PRL_A} with the strength $t^{(2)} \propto t_2 \Delta_M/E_F$  as a result of two subsequent tunneling events, see Fig. \ref{nu2}. 
As follows directly from considering all possible second-order processes in leading order in $\Delta_M$,  there are now two uncoupled modes at each edge: $L_{1  1\bar 1}$, $L_{1  1  1}$ (lower edge) and $R_{N \bar 1\bar1}$, $R_{N \bar 1 1}$   (upper edge).


\begin{thebibliography}{99}




\bibitem{review_AHE} N. Nagaosa,
J. Sinova, S. Onoda, A. H. MacDonald, and N. P. Ong,
Rev. Mod. Phys. {\bf 82}, 1539 (2010).

\bibitem{review_AHE_TI} K. He, Y. Wang, and Q.-K. Xue, Natl. Sci. Rev. {\bf 1}, 39, (2014).


\bibitem{Klitzing} K. v. Klitzing, G. Dorda, and M. Pepper, Phys. Rev. Lett. {\bf 45}, 494 (1980).

\bibitem{Tsui} D. C. Tsui, H. L. Stormer, and A. C. Gossard, Phys. Rev. Lett. {\bf 48}, 1559 (1982).

\bibitem{Pankratov} O.A. Pankratov, Phys. Lett. A {\bf 121}, 360 (1987).


\bibitem{Haldane} F. D. M. Haldane, Phys. Rev. Lett. {\bf 61}, 2015 (1988).

\bibitem{Zhang}  C.-X. Liu, X.-L. Qi, X. Dai, Z. Fang, and S.-C. Zhang,
Phys. Rev. Lett. {\bf 101}, 146802 (2008).

\bibitem{Zhang_2} R. Yu,
W. Zhang, H.-J. Zhang, S.-C. Zhang, X. Dai, and Z. Fang, 
Science {\bf 329}, 61 (2010).


\bibitem{Wang_2013}
Z. F. Wang, Z. Liu, and F. Liu, Phys. Rev. Lett. {\bf 110}, 196801 (2013).



\bibitem{Qiao} Z. H. Qiao {\it et al.}, 
Phys. Rev. B {\bf 82}, 161414(R) (2010).


\bibitem{Mokrousov_2012} H. B. Zhang, C. Lazo, S. Blugel, S. Heinze, and Y. Mokrousov, 
Phys. Rev. Lett. {\bf 108}, 056802 (2012).



\bibitem{Niu_2014}  Z. Qiao  {\it et al.},
Phys. Rev. Lett. {\bf 112}, 116404 (2014).


\bibitem{Ezawa_2012}
M. Ezawa, Phys. Rev. Lett. {\bf 109}, 055502 (2012).

\bibitem{Garrity_2013}
K. F. Garrity and D. Vanderbilt, Phys. Rev. Lett. {\bf 110}, 116802 (2013).


\bibitem{Kagome_2011} Z.-Y. Zhang, J. Phys. Condens. Matter 23, 365801 (2011).


\bibitem{Wu_2008} C. Wu, Phys. Rev. Lett. {\bf 101}, 186807 (2008).



\bibitem{Zhang_3} C. Chang
{\it et al.}, Science {\bf 340}, 167 (2013).

\bibitem{ETHZ} G. Jotzu, M. Messer, R. Desbuquois, M. Lebrat, T. Uehlinger, D. Greif, and T. Esslinger, Nature {\bf 515}, 237 (2014).





\bibitem{Lebed} A. G. Lebed, JETP Lett. {\bf 43}, 174 (1986).

\bibitem{Montambaux} D. Poilblanc, 
G. Montambaux, M. Heritier, and P. Lederer, 
Phys. Rev. Lett. {\bf 58}, 270 (1987).

\bibitem{Lebed_Gorkov} L. P. Gor'kov and A. G. Lebed, Phys. Rev. B {\bf 51}, 3285 (1995).

\bibitem{Yakovenko_PRB} V. M. Yakovenko, Phys. Rev. B {\bf 43}, 11353 (1991).

\bibitem{Wegsheider} R. A. Deutschmann,
W. Wegscheider, M. Rother, M. Bichler, G. Abstreiter, C. Albrecht, and J. H. Smet,
 Phys. Rev. Lett. {\bf 86}, 1857 (2001).

\bibitem{Kane_PRL} C. L. Kane, R. Mukhopadhyay, and T. C. Lubensky, Phys. Rev. Lett. {\bf 88}, 036401 (2002).


\bibitem{Stripes_PRL} J. Klinovaja and D. Loss, Phys. Rev. Lett. {\bf 111}, 196401 (2013).

\bibitem{Kane_PRB} J. C. Y. Teo and C. L. Kane, Phys. Rev. B {\bf 89}, 085101 (2014).

\bibitem{Stripes_arxiv} J. Klinovaja and D. Loss, Eur. Phys. J. B {\bf 87}, 171 (2014).

\bibitem{Stripes_nuclear} T. Meng,
P. Stano, J. Klinovaja, and D. Loss, 
Eur. Phys. J. B {\bf 87}, 203 (2014). 


\bibitem{Stripe_PRL_exp} K. Kobayashi,
H. Satsukawa, J. Yamada, T. Terashima, and S. Uji, 
Phys. Rev. Lett. {\bf 112}, 116805 (2014).

\bibitem{yaroslav} J. Klinovaja and Y. Tserkovnyak, Phys. Rev. B {\bf 90}, 115426 (2014). 

\bibitem{Neupert} T. Neupert,
C. Chamon, C. Mudry, and R. Thomale, 
Phys. Rev. B {\bf 90}, 205101 (2014).

\bibitem{Oreg} E. Sagi and Y. Oreg, arXiv:1403.1791.



\bibitem{braunecker_prb_10} B. Braunecker,
G. I. Japaridze, J. Klinovaja, and D. Loss,
 Phys. Rev. B \textbf{82}, 045127 (2010).

\bibitem{nanomagnets} B. Karmakar {\it et al.}, Phys. Rev. Lett. {\bf 107}, 236804 (2011).

\bibitem{Two_field_Klinovaja_Stano_Loss_2012} J. Klinovaja, P. Stano, and D. Loss, Phys. Rev. Lett. {\bf 109}, 236801 (2012).

\bibitem{Flensberg} M. Kjaergaard, K. Wolms, and K. Flensberg, Phys. Rev. B {\bf 85}, 020503 (2012).


\bibitem{helical_graphene} J. Klinovaja and D. Loss, Phys. Rev. X {\bf 3}, 011008 (2013).






\bibitem{skyrmion_2} A. Bogdanov and A. Hubert, J. Magn. Magn. Mater. {\bf 138}, 255 (1994).


\bibitem{skyrmion_1} S. Muhlbauer 
B. Binz, F. Jonietz, C. Pfleiderer, A. Rosch, A. Neubauer, R. Georgii, and P. Boni,
 Science, {\bf 323} 915 (2009).

\bibitem{skyrmion_4} X. Z. Yu, 
	 Y. Onose,	 N. Kanazawa,	 J. H. Park,	 J. H. Han,	 Y. Matsui,	 N. Nagaosa,	 and Y. Tokura,
	  Nature {\bf 465}, 901 (2010).

\bibitem{skyrmion_tokura} Y. Li {\it et al.},
 Phys. Rev. Lett. {\bf 110}, 117202  (2013).

\bibitem{Tserkovnyak_2014} Y. Tserkovnyak, D.~A. Pesin, and D. Loss, arXiv:1411.2070.



\bibitem{Daniel_RKKY} B. Braunecker, P. Simon, and D. Loss, Phys. Rev. B {\bf 80}, 165119 (2009).

\bibitem{Exp_RKKY} C. P. Scheller,
T.-M. Liu, G. Barak, A. Yacoby, L. N. Pfeiffer, K. W. West, and D. M. Zumbuhl,
 Phys. Rev. Lett. {\bf 112}, 066801 (2014).

\bibitem{RKKY_Basel} J. Klinovaja,
P. Stano, A. Yazdani, and D. Loss,
  Phys. Rev. Lett. {\bf 111}, 186805 (2013).

\bibitem{RKKY_Simon} B. Braunecker and P. Simon, Phys. Rev. Lett. {\bf 111}, 147202 (2013).

\bibitem{RKKY_Franz} M. Vazifeh and M. Franz,  Phys. Rev. Lett. {\bf 111}, 206802 (2013).


\bibitem{MF_PRB} J. Klinovaja and D. Loss, Phys. Rev. B {\bf 86}, 085408 (2012).


\bibitem{Appendix} See Supplemental Material at .. for the topological phase diagram, edge modes wavefunctions, and IQAHE at higher values of $l$.



\bibitem{giamarchi_book} T. Giamarchi, {\it Quantum Physics in One Dimension} (Oxford University Press, New York, 2003).


\bibitem{cheng_PF} M. Cheng, Phys. Rev. B {\bf 86}, 195126 (2012).

\bibitem{Stern_PF}  N. H. Lindner,
E. Berg, G. Refael, and A. Stern, 
Phys. Rev. X
{\bf 2}, 041002 (2012).


\bibitem{David} D. Clarke, J. Alicea, and K. Shtengel, Nat. Commun. {\bf 4}, 1348
(2013).

\bibitem{Mong} R. S. K. Mong {\it et al.},
 Phys. Rev. X {\bf 4}, 011036 (2014).
 
\bibitem{PF_TI} J. Klinovaja, A. Yacoby, and D. Loss, Phys. Rev. B {\bf 90}, 115426 (2014).


\bibitem{Barkeshli} M. Barkeshli, C.-M. Jian, and X.-L. Qi, Phys. Rev. B {\bf 87}, 045130 (2013).

\bibitem{Vaezi} A. Vaezi, Phys. Rev. B {\bf 87}, 035132 (2013).

\bibitem{PF_1} J. Klinovaja and D. Loss, Phys. Rev. Lett. {\bf 112}, 246403 (2014).

\bibitem{PF_2} J. Klinovaja and D. Loss, Phys. Rev. B {\bf 90}, 045118 (2014).

\bibitem{Vaezi_2} A. Vaezi and M. Barkeshli, arXiv:1403.3383.



\end{thebibliography}

\begin{thebibliography}{99}

\bibitem{Stripes_PRL_A} J. Klinovaja and D. Loss, Phys. Rev. Lett. 111, 196401 (2013).

\bibitem{Stripes_arxiv_A} J. Klinovaja and D. Loss, Eur. Phys. J. B {\bf 87}, 171 (2014).

\bibitem{giamarchi_book_A} T. Giamarchi, {\it Quantum Physics in One Dimension} (Oxford University Press, New York, 2003).

\bibitem{MF_PRB_A} J. Klinovaja and D. Loss, Phys. Rev. B {\bf 86}, 085408 (2012).

\bibitem{Two_field_Klinovaja_Stano_Loss_2012_A} J. Klinovaja, P. Stano, and D. Loss, Phys. Rev. Lett. {\bf 109}, 236801 (2012).


\end{thebibliography}
\end{document}